\documentclass[12pt,preprint]{aastex}

\slugcomment{Submitted to \apjl}

\shorttitle{Inflow motion in massive cores} \shortauthors{Wu et al.}

\begin{document}

\title{Signatures of inflow motion in cores of massive star
formation:\\
Potential collapse candidates}

\author{Yuefang Wu\altaffilmark{1},
        Christian Henkel\altaffilmark{2},
        Rui Xue\altaffilmark{1,3},
        Xin Guan\altaffilmark{1},
        Martin Miller\altaffilmark{4}}

\altaffiltext{1}{Department of Astronomy, Peking Univ., 100871
Beijing, China, yfwu@bac.pku.edu.cn}
\altaffiltext{2}{Max-Planck-Institut f{\"u}r Radioastronomie, Auf
                 dem H{\"u}gel 69, 53121 Bonn, Germany}
\altaffiltext{3}{Present affiliation: National Astronomical Observatory,
                 Chinese Academy of Sciences, 100012 Beijing, China}
\altaffiltext{4}{I. Physikalisches Institut der Universit{\"a}t zu K{\"o}ln,
                 Z{\"u}lpicher Stra{\ss}e 77, 50937, K{\"o}ln, Germany}

\begin{abstract}
Using the IRAM 30 m telescope, a mapping survey in optically thick
and thin lines was performed towards 46 high mass star-forming
regions. The sample includes UC~H{\sc ii} precursors and UC~H{\sc ii}
regions. Seventeen sources are found to show "blue profiles", the
expected signature of collapsing cores. The excess of sources with
blue over red profiles ([$N_{\rm blue}$ -- $N_{\rm red}$]/$N_{\rm total}$)
is 29\% in the HCO$^+$ $J$=1--0 line, with a probability of 0.6\%
that this is caused by random fluctuations. UC~H{\sc ii} regions show
a higher excess (58\%) than UC~H{\sc ii} precursors (17\%), indicating
that material is still accreted after the onset of the UC~H{\sc ii}
phase. Similar differences in the excess of blue profiles as a function
of evolutionary state are not observed in low mass star-forming
regions. Thus, if confirmed for high mass star-forming sites, this
would point at a fundamental difference between low- and high-mass
star formation. Possible explanations are inadequate thermalization,
stronger influence of outflows in massive early cores, larger gas
reserves around massive stellar objects or different trigger
mechanisms between low- and high- mass star formation.

\end{abstract}


\section{Introduction}

Inflow motion is a fundamental phenomenon during stellar formation.
Although the search for inflow is usually more difficult than that
for outflow, studies of inflow have made great progress since the
1990s. In low-mass star forming regions, inflow motions were
detected at different evolutionary stages, including Class --I,
Class 0 and Class I cores \citep{zho93, mmt97, lee99, gre00,
evans03}. Recently, a number of inflow candidates were found in high
mass star formation regions. Among a sample of 28 massive cores, 12
were found to show line profiles that peak at blue-shifted
velocities (hereafter "blue profiles"; see Sect.\,3.1), the expected
signature of inflow \citep{we03}. \citet{fws05} (hereafter FWS05)
detected such asymmetric profiles in 22 cores within a sample of 77
high-mass proto-stellar objects (HMPOs). Most recently,
\citet{wyr06} detected 9 sources with a blue profile in a sample of
12 ultracompact (UC)~H{\sc ii} regions.

Variation of inflow motion with time is critical for high mass star
formation. It has been indicated that when a protostar reaches $>$10
M$_{\odot}$ it can generate enough radiation pressure to halt
spherical infall and inhibit its mass increasing\citep{wc87}.
Observationally, however, it is not yet clear how inflow is related
to the evolution of massive (proto)stars.  To study this problem, we
have carried out a survey for a sample including both cores of
UC~H{\sc ii} regions and precursors of UC~H{\sc ii} regions.

While previous surveys using single point observations provided some
statistical evidence for the occurrence of infall within massive
cores, blue profiles can also be caused by rotation. Therefore maps
of the molecular environment are indispensable. Mapping also allows
us to locate the center of the inflow and to identify cores that are
simultaneously showing evidence for in- and outflow.

Therefore, we conducted a mapping survey including 46 high mass
star-forming regions which were selected applying three criteria:
(1) The sources must have been mapped in the submillimeter or
millimeter wavelengths with continuum or spectroscopy; (2)
signal-to-noise ratios should be $>$5 at 350\,$\mu$m (Mueller et al.
2002) and higher at other wavelengths; (3) there should be no other
core within one arcmin \citep{zhs97, hnb98, tie98, hat00, mbc00,
bsp02, mse02}. With respect to their stellar content, we can divide
the sample into two different groups of targets: (1) Thirty three
sources lack 6 cm continuum emission and are precursors of UC~H{\sc
ii} regions or HMPOs \citep{mbc00, bsp02}. Among these, thirty are
hosting a luminous IRAS source. The remaining three are associated
with IRAC (the InfraRed Array Camera on the Spitzer Space Telescope)
point sources (W3-W and W3-SE) or are not hosting an IRAC source
(18454--3). All 33 cores comprise `group~I'. (2) Thirteen UC~H{\sc
ii} regions are assigned to `group~II'. This letter presents  a list
of the identified collapse candidates and provides the statistics of
blue excesses. Detailed properties of individual cores will be
analyzed in a future paper.

\section{Observations}

The observations were performed with the IRAM 30\,m telescope at
Pico Veleta, Spain, from July 28 to Aug. 1, 2005. Four receivers
were used simultaneously, usually two at $\lambda$$\sim$3\,mm and
two at $\lambda$$\sim$1.3\,mm (A/B configuration). For some sources,
none of the four 3 and 1.3\,mm lines were optically thin. In these
cases the tracer lines were changed employing two receivers at
$\lambda$$\sim$2\,mm and the other two to cover the upper part of
the 1.3\,mm window (C/D configuration). The lines and corresponding
beam sizes, efficiencies, and channel widths
are given in Table 1. The channel spacing and the bandwidth are
78.125 kHz and 105 MHz respectively. The weather was extremely good
for summer conditions, allowing us to observe the HCO$^+$ $J$=3--2
transition at 268\,GHz and leading to 3 and 1.3\,mm (2 and 1.2\,mm)
system temperatures of order 150 and 400\,K (200 and 550\,K) on a
$T_{\rm A}^*$ scale. Pointing and calibration were checked by
continuum measurements of the standard sources W3(OH), G34.24, and
NGC 7027 and were found to be better than 4\arcsec\ and $\pm$20\%,
respectively. All observations were carried out in a position
switching mode. For each source we observed a nine point map in a
cross pattern with a spacing of 15\arcsec. If inflow signature was
detected, the map was enlarged in most cases to cover the entire
region showing this signature. The on-source integration time per
position was 1 minute, yielding a $T_{\rm A}$* 1$\sigma$ noise level
of 0.07\,K for the 3\,mm N$_2$H$^+$ (1--0) line. For the data
analysis, the GILDAS software package (CLASS/GREG) was used
\citep{gl00}.

\section{Results and discussion}

\subsection{Blue profile identification}

For self-absorbed optically thick lines, the classical signature of
inflow is a double peaked profile with the blue-shifted peak being
stronger, or a line asymmetry with the peak skewed to the blue side.
While optically thin lines should show a single velocity component
peaking at the line center.

Among the 46 cores observed, five (05490+2658, G31.41+0.31, 18454-3,
18454-4, 19266+1745) will be ignored because they show either too
complex spectral profiles, inhibiting a detailed analysis, or a lack
of optically thin lines. Estimates of optical depths were obtained
from line ratios between different isotopomers of CO and CS and from
the relative intensities of individual hyperfine components in the
case of C$^{17}$O and N$_2$H$^+$. C$^{18}$O, C$^{17}$O, C$^{34}$S
and N$_2$H$^+$ tend to be optically thin, while CS is optically
thick. HCO$^+$ opacities could not be estimated. However, the
similarity of HCO$^+$ and CS line shapes (see Sect.\,3.2) as well as
the results of \cite{gre00} and FWS05 clearly indicate that HCO$^+$
is also optically thick.

The 41 remaining sources were detected in at least one optically
thick and one optically thin line. A blue profile caused by inflow
motion with velocity $v \propto r^{-1/2}$ in a region with higher
excitation temperature ($T_{ex}$) inside requires $T_{\rm
A}$*(B)/$T_{\rm A}$*(R) $>$ 1. Here $r$ is the radius of the
collapsing core \citep{zho93}. $T_{\rm A}$*(B) and $T_{\rm A}$*(R)
are the blue and red peak intensities of the optically thick line.
We also define a dimensionless asymmetry parameter following
\cite{mmt97}, $\delta V$ = ($V_{\rm thick}$-$V_{\rm
thin}$)/$\Delta$$V_{\rm thin}$. $V_{\rm thick}$ is the peak velocity
of the opaque line, $V_{\rm thin}$ and $\Delta$$V_{\rm thin}$ denote
the peak velocity and width of the optically thin line. Only for
$\delta V < -0.25$ or $> 0.25$ the line profile is rated blue or
red, respectively.

Our sources (Table~2) discriminate among five main types of line
shapes: (1) cores with lines showing a ``blue profile'' (in the
following denoted with B); (2) cores with lines showing a "red
profile" (R); (3) cores exhibiting blue and red profiles at
different spatial positions (BRS); (4) cores where some lines show a
blue profile, while others display a red profile (BRL); (5) cores
without obvious asymmetric lines (S). Only cores showing at least
one line of type B, but no lines of type R are identified as targets
potentially undergoing inflow motion.

\subsection{Collapse candidates and their profile ``excess''} \label{bozomath}

With the criteria outlined in Sect.\,3.1, seventeen inflow
candidates are identified (see Table~2). Ten belong to group~I and
seven are part of group~II. To provide a typical example, Fig.\,1
shows the infall signature of the group~I core W3-SE. Fig. 1a
displays the HCO$^+$\,(1--0) spectra, showing the angular size of
the core. Fig. 1b shows a number of profiles towards the central
position. The HCO$^+$\,(1--0) and (3--2) lines as well as the
CS\,(3--2) transition show the blue asymmetry. For the
HCO$^+$\,(1--0) line this is also demonstrated in the
position-velocity (P-V) diagram of Fig.\,1c. For comparison,
Fig.\,1d shows a P-V diagram of the optically thin C$^{18}$O\,(1--0)
emission.

The quantity ``excess'' as defined by \cite{mmt97} is $E$ = ($N_{\rm
B} - N_{\rm R}$)/$N_{\rm T}$, where $N_{\rm B}$ and $N_{\rm R}$ mark
the numbers of sources with blue and red profiles. $N_{\rm T}$ is
the total number of sources. For our survey the excess was
calculated for the two HCO$^+$ transitions and the CS\,(3--2) line.
Fig. 2 shows the log[$T_{\rm A}$*(B)/$T_{\rm A}$*(R)] and $\delta$V
(see Sect.\,3.1) distributions of the three individual lines.
Statistical results are given in Table~3. The observed excess
derived from the HCO$^+$\,(1--0) and (3--2) lines is 0.29 and 0.11,
respectively. Both are larger than those obtained by FWS05 for the
same lines (0.15 and 0.04). For the CS transition we obtain 0.29. To
evaluate the statistical significance of the determined values, we
conducted the binomial test (see FWS05 and references therein).
Probabilities that the excesses are a product of a random
distribution are given in the last column of Table~3. These are
0.006 and 0.01 for HCO$^+$\,(1--0) and CS (3--2) respectively.
Apparently, both lines are sensitive tracers of potential inflow
motion in massive cores.

To evaluate differences between the two classes of cores (I and II;
see Sect.\,1) with respect to the excess, we used the HCO$^+$\,(1--0)
line, which was mapped in the largest number of sources. The results
listed in the lower part of Table~3 include 16 sources with profiles
of type B. The excesses observed for group~I and II are 0.17 and 0.58,
respectively.

Twenty of our 46 sources overlap with those of FWS05. Among them are
19 group~I sources (out of 33), but only one source is from group~II
(out of 13). Our study includes various CO and CS lines. We also
made maps. Thus we can view the common objects from a different
perspective and can check, how far the choice of different molecular
transitions and the presence of maps is leading to contradictions
with previously published results. Differences are indeed
significant. For eight of the 19 overlapping type I cores we obtain
different line asymmetry classification, emphasizing the need for
detailed maps. Nevertheless, the overall difference in the
HCO$^+$\,(1--0) excess is negligible (0.17 versus 0.15).

To summarize, both data sets indicate that the HCO$^+$\,(1--0)
excess is low for UC~H{\sc ii} precursors. For  UC~H{\sc ii}
regions, our results and those of \cite{wyr06} suggest that the
excess is larger and more significant. From the binomial test for
group~I and II, the probability that the blue excesses (0.17 and
0.58 respectively) are arising by chance is 0.13 and 0.008,
respectively.

\subsection{A comparison with low mass star-forming surveys}

While low mass star-forming regions show infall from the Class --I
to the Class I stages of evolution, high mass star-forming regions
also exhibit infall signatures from their earliest stages till a UC
H{\sc ii} region has formed \citep{wzx05,bkl06,qzm06}. In low mass
cores the profile excess was found to be 0.30, 0.31 and 0.31 for
Class --I, 0 and I core samples in the HCO$^+$\,(3--2) line
(\cite{evans03} and references therein). There seem to be no
significant differences among the cores in different evolutionary
phases. However, our samples show the excess of UC~H{\sc ii} regions
far surpassing that of the UC~H{\sc ii} precursors. This may point
to fundamental differences between low and high mass star-forming
conditions. Possible causes to the higher blue excess in Group II
sources may be: (1) The molecular gas surrounding UC H{\sc ii}
regions may be more adequately thermalized to show the blue excess,
i.e. the excitation temperature of specific lines may increase more
monotonically towards the center. Thus all lines may produce blue
profiles indicating infall motion, while in younger cores still some
lines may show red profiles. (2) The amount of dense cool gas is
larger towards younger objects. Outflows of dense molecular gas may
be more active around Group I objects, shaping more red profiles.
(3) Low mass cores are relatively isolated and their gas supply is
limited. Simulations showed that this may halt the increase of
inflow \citep{vor05}. However, high mass stars form in giant
molecular clouds and their inflow motions are not easily halted by
the exhaustion of molecular gas before most of it is dispelled. (4)
In low mass cores, star formation may be spontaneous. In high mass
cores, collapse may be trigged by extrinsic disturbances and the
collapse may take more time to develop.

With respect to potential selection effects, we used the same
criteria to identify the targets of the two separate groups of
sources. Since this study is based on a limited number of sources,
more data quantifying the blue excess as a function of evolutionary
stage would be highly desirable.

\section{Outlook}

We have carried out a mapping survey towards 46 molecular cores
associated with massive star formation. Seventeen collapse
candidates were identified. Among them are 10 UC H{\sc ii}
precursors and 7 UC~H{\sc ii} regions. Overall, statistical results
indicate a predominance of blue over red profiles which is
surprisingly similar to that obtained towards cores forming low mass
stars. Among high mass star-forming sites, the probability to detect
blue profiles seems to depend on evolutionary stage and increases
from UC~H{\sc ii} precursors to UC~H{\sc ii} regions. Toward low
mass star-forming sites, however, this effect is not observed,
suggesting a more fundamental difference in the way stellar masses
are assembled. Larger line surveys and more detailed maps in various
molecular transitions are needed to improve statistical evidence in
order to confirm or to reject this potentially important finding.

\acknowledgments

We are grateful to the IRAM staff for their assistance and F.
Wyrowski for useful discussions. This research is supported by the
Grant 10128306 and 10733030 of NSFC.

\clearpage

\begin{figure}
\epsscale{.50} \plotone{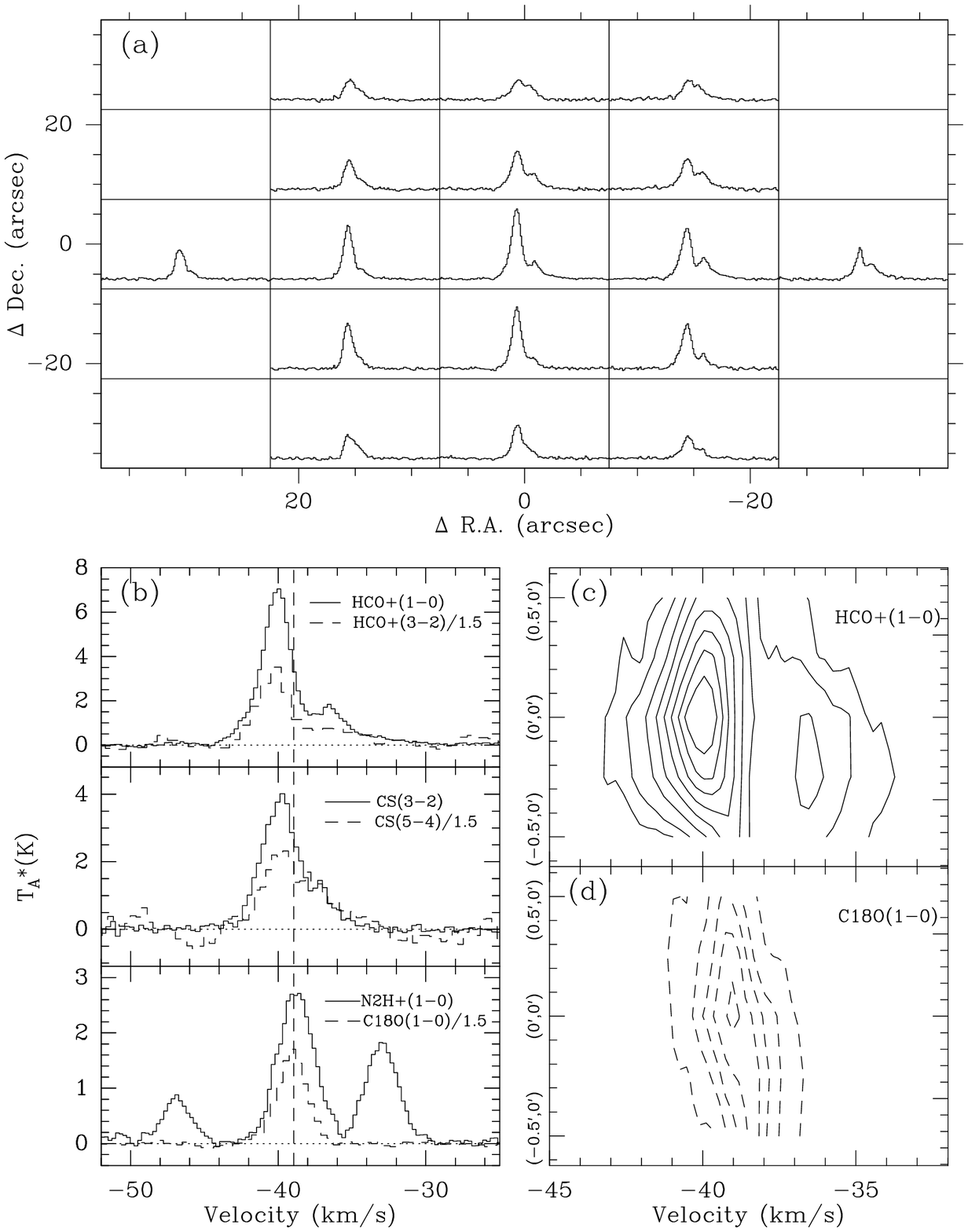} \caption{W3-SE. {\bf a.} HCO$^+$
(1--0) grid. The X and Y axes denote R.A. and Dec. offsets in
arcseconds relative to the NH$_3$ peak position of \cite{tie98}.
{\bf b.} Spectra towards the central position. Molecular species and
transition are given at the upper-right corner of each inset. {\bf
c.} Position-velocity diagram of the optically thick HCO$^+$\,(1--0)
line along $\Delta$Dec. = 0. Contour levels are 0.5\,K (3$\sigma$)
and 1.0 to 5.8\,K by 0.8\,K. {\bf d.} Position-velocity diagram of
the optically thin C$^{18}$O\,(1--0) line along $\Delta$Dec. = 0.
Contour levels are 0.3\,K (3$\sigma$) and
 1.0 to 2.5\,K by 0.5\,K.
\label{fig1}}
\end{figure}

\clearpage

\begin{figure}
\epsscale{.60} \plotone{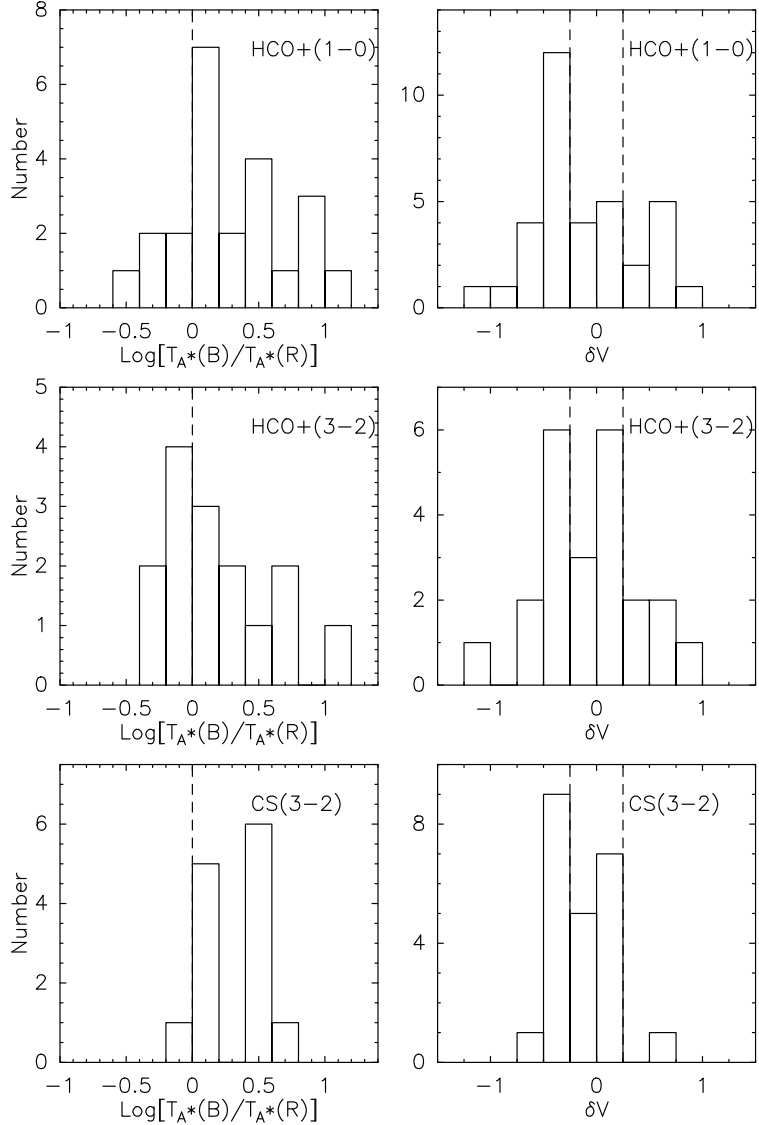} \caption{The distribution of
log\,[$T_{\rm A}^*$(B)/$T_{\rm A}^*$(R)] (left; ratio of blue versus
red peak intensity) and $\delta$V (right; the line asymmetry
parameter, see Sect.\,3.1). The upper, middle and lower panels show
results from HCO$^+$(1--0), HCO$^+$(3--2) and CS(3--2),
respectively. There is a total of 41 cores observed in the
HCO$^+$(1--0), 28 in the HCO$^+$(3--2), and 28 in the CS(3--2) line.
For the $\delta V$ histograms, BRS sources (for a definition, see
Sect. 3.1) had to be excluded (6, 5, and 5 in the three transitions,
respectively). There are additional 12, 8, and 10 cores, for which no
blue versus red peak intensity ratio could be obtained.}

\end{figure}

\clearpage

\begin{deluxetable}{lllllll}
\tabletypesize{\small} \tablewidth{0pt} \tablecolumns{6}
\tablecaption{Observed lines$^{\rm a)}$\label{tb1}}

\startdata \tableline\tableline
No. &   Line                &   Frequency   &   HPBW    &   $\eta_{\rm mb}$ &   $\Delta V_{\rm res}$ \\
    &                       &    (GHz)      & (arcsec)  &                   &    (km\,s$^{-1}$)   \\
\tableline
1   &   $HCO^{+} (1-0)$      &    89.18852   &   27.6    &   0.77        &   0.263 \\
2   &   $HCO^{+} (3-2)$      &   267.55763   &    9.2    &   0.45        &   0.112 \\
3   &   $CS (3-2)$           &   146.96905   &   16.7    &   0.69        &   0.159 \\
4   &   $CS (5-4)$           &   244.93561   &   10.0    &   0.49        &   0.122 \\
5   &   $N_{2}H^{+} (1-0)$   &    93.17378   &   26.4    &   0.77        &   0.251 \\
6   &   $C^{18}O (1-0)$      &   109.78218   &   22.4    &   0.75        &   0.213 \\
7   &   $C^{18}O (2-1)$      &   219.56033   &   11.2    &   0.55        &   0.137 \\
8   &   $C^{17}O (1-0)$      &   112.35928   &   21.9    &   0.74        &   0.209 \\
9   &   $C^{17}O (2-1)$      &   224.71437   &   10.9    &   0.54        &   0.133 \\
10   &   $C^{34}S (5-4)$      &   241.01618   &   10.2    &   0.50        &   0.122 \\

\enddata

\bigskip
\tablenotetext{\ }{a) HPBW: half power beamwith; $\eta_{\rm mb}$:
beam efficiency; $\Delta V_{\rm res}$: channel width }
\end{deluxetable}

\clearpage

\begin{deluxetable}{lcccrc|lcccrc}
\tablecaption{Profiles of the observed sources\label{tb2}}
\tabletypesize{\tiny} \tablewidth{0pt} \tablecolumns{9} \startdata
\tableline\tableline Source   &  \hskip -0.2truecm $\alpha$$^b$ &
\hskip -0.2truecm $\delta$$^b$ &  \hskip -0.3truecm $D^{c}$ & \hskip
-0.3truecm & \hskip -0.3truecm & Source & \hskip -0.2truecm
$\alpha$$^b$ & \hskip -0.2truecm $\delta$$^b$ & \hskip -0.3truecm
$D^{c}$ & \hskip
-0.3truecm  & \hskip -0.3truecm  \\
Name$^{a}$     &  \hskip -0.2truecm J2000 & \hskip -0.2truecm J2000
& \hskip -0.3truecm (kpc) & \hskip -0.3truecm Profile$^{d}$ & \hskip
-0.3truecm Ref. & Name$^{a}$  &  \hskip -0.2truecm J2000 & \hskip
-0.2truecm J2000 &
\hskip -0.3truecm (kpc) & \hskip -0.3truecm Profile$^{d}$ & \hskip -0.3truecm Ref. \\
\tableline

W3-W$^I$          & \hskip -0.2truecm $02~25~32.4$ & \hskip -0.2truecm $+62~06~01$ & \hskip -0.2truecm $1.95$ & \hskip -0.3truecm $B$ & \hskip -0.2truecm 1                   & 18488+0000SE$^I$ & \hskip -0.2truecm $18~51~25.6$ & \hskip -0.2truecm $+00~04~07$ & \hskip -0.2truecm $5.4$ & \hskip -0.3truecm $BRL$  & \hskip -0.2truecm 3,7 \\
W3-C$^{II}$       & \hskip -0.2truecm $02~25~39.5$ & \hskip -0.2truecm $+62~05~51$ & \hskip -0.2truecm $2.3$ & \hskip -0.3truecm $BRS$ & \hskip -0.2truecm 1                        & G34.26+0.15$^{II}$ & \hskip -0.2truecm $18~53~18.4$ & \hskip -0.2truecm $+01~14~56$ & \hskip -0.2truecm $3.7$ & \hskip -0.3truecm $In,B$  & \hskip -0.2truecm 6 \\
W3-SE$^I$         & \hskip -0.2truecm $02~25~54.5$ & \hskip -0.2truecm $+62~04~11$ & \hskip -0.2truecm $2.3$ & \hskip -0.3truecm B & \hskip -0.2truecm 1                          & 18521+0134$^I$    & \hskip -0.2truecm $18~54~40.8$ & \hskip -0.2truecm $+01~38~02$ & \hskip -0.2truecm $5.0$ & \hskip -0.3truecm $B$ &   \hskip -0.2truecm 3,7 \\
05358+3543$^I$    & \hskip -0.2truecm $05~39~10.4$ & \hskip -0.2truecm $+35~45~19$ & \hskip -0.2truecm $1.8$ & \hskip -0.3truecm $BRS$ & \hskip -0.2truecm 3,7           & 18530+0215$^I$    & \hskip -0.2truecm $18~55~34.2$ & \hskip -0.2truecm $+02~19~08$ & \hskip -0.2truecm $5.1$ & \hskip -0.3truecm $S$  & \hskip -0.2truecm 3,7 \\
05490+2658$^I$    & \hskip -0.2truecm $05~52~12.9$ & \hskip -0.2truecm $+26~59~33$ & \hskip -0.2truecm $2.1$ & \hskip -0.3truecm $ ... $ & \hskip -0.2truecm 3,7             & S76E$^{II}$      & \hskip -0.2truecm $18~56~11.0$ & \hskip -0.2truecm $+07~53~28$ & \hskip -0.2truecm $2.1$ & \hskip -0.3truecm $S?$   & \hskip -0.2truecm 4,9\\
G10.47+0.03$^{II}$ & \hskip -0.2truecm $18~08~38.2$ & \hskip -0.2truecm $-19~51~50$ & \hskip -0.2truecm $5.8$ & \hskip -0.3truecm $B$ & \hskip -0.2truecm 10                        & 18553+0414NE$^I$  & \hskip -0.2truecm $18~57~53.4$ & \hskip -0.2truecm $+04~18~15$ & \hskip -0.2truecm $0.6$ & \hskip -0.3truecm $B$  & \hskip -0.2truecm 3,7 \\
G12.42+0.50$^{II}$ & \hskip -0.2truecm $18~10~51.8$ & \hskip -0.2truecm $-17~55~56$ & \hskip -0.2truecm $2.1$ & \hskip -0.3truecm $B$ & \hskip -0.2truecm 2                     & 19012+0536$^I$    & \hskip -0.2truecm $19~03~45.1$ & \hskip -0.2truecm $+05~40~40$ & \hskip -0.2truecm $4.6$ & \hskip -0.3truecm $B$   & \hskip -0.2truecm 3,7 \\
G12.89+0.49$^{II}$ & \hskip -0.2truecm $18~11~51.3$ & \hskip -0.2truecm $-17~31~29$ & \hskip -0.2truecm $3.5$ & \hskip -0.3truecm $BRL$ & \hskip -0.2truecm 2,3,9                     & 19092+0841SW$^I$ & \hskip -0.2truecm $19~11~36.7$ & \hskip -0.2truecm $+08~46~20$ & \hskip -0.2truecm $4.48$ & \hskip -0.3truecm $BRL$   & \hskip -0.2truecm 5 \\
G13.87+0.28$^{II}$ & \hskip -0.2truecm $18~14~35.4$ & \hskip -0.2truecm $-16~45~37$ & \hskip -0.2truecm $4.4$ & \hskip -0.3truecm $S?$  & \hskip -0.2truecm 10                     & G43.89-0.79$^{II}$  & \hskip -0.2truecm $19~14~26.2$ & \hskip -0.2truecm $+09~22~34$ & \hskip -0.2truecm $4.2$ & \hskip -0.3truecm $B$   & \hskip -0.2truecm 10 \\
18144-1723NW$^I$  & \hskip -0.2truecm $18~17~23.8$ & \hskip -0.2truecm $-17~22~09$ & \hskip -0.2truecm $4.33$ & \hskip -0.3truecm $R$   & \hskip -0.2truecm 5                       & 19217+1651N$^I$   & \hskip -0.2truecm $19~23~58.8$ & \hskip -0.2truecm $+16~57~45$ & \hskip -0.2truecm $10.5$ & \hskip -0.3truecm $B$   & \hskip -0.2truecm 3,7 \\
18182-1433$^I$    & \hskip -0.2truecm $18~21~07.9$ & \hskip -0.2truecm $-14~31~53$ & \hskip -0.2truecm $4.5$ & \hskip -0.3truecm $B$   & \hskip -0.2truecm 3,7                      & 19266+1745$^I$    & \hskip -0.2truecm $19~28~54.0$ & \hskip -0.2truecm $+17~51~56$ & \hskip -0.2truecm $0.3$ & \hskip -0.3truecm $ ... $   & \hskip -0.2truecm 3,7 \\
G19.61$^{II}$     & \hskip -0.2truecm $18~27~37.9$ & \hskip -0.2truecm $-11~56~07$ & \hskip -0.2truecm $4.0$ & \hskip -0.3truecm $S?$   & \hskip -0.2truecm 2                       & 19410+2336$^I$    & \hskip -0.2truecm $19~43~11.4$ & \hskip -0.2truecm $+23~44~06$ & \hskip -0.2truecm $2.1$ & \hskip -0.3truecm $B$   & \hskip -0.2truecm 3,7 \\
18264-1152$^I$    & \hskip -0.2truecm $18~29~14.3$ & \hskip -0.2truecm $-11~50~26$ & \hskip -0.2truecm $3.5$ & \hskip -0.3truecm $R$   & \hskip -0.2truecm 3,7                       & S86SE$^I$     & \hskip -0.2truecm $19~43~49.7$ & \hskip -0.2truecm $+23~28~41$ & \hskip -0.2truecm $1.9$ & \hskip -0.3truecm $BRS$   & \hskip -0.2truecm 4 \\
18306-0835$^I$    & \hskip -0.2truecm $18~33~21.8$ & \hskip -0.2truecm $-08~33~38$ & \hskip -0.2truecm $4.9$ & \hskip -0.3truecm $R$   & \hskip -0.2truecm 3,7                       & S87N$^I$          & \hskip -0.2truecm $19~46~20.6$ & \hskip -0.2truecm $+24~36~04$ & \hskip -0.2truecm $2.3$ & \hskip -0.3truecm $S?$   & \hskip -0.2truecm 4 \\
G24.49-0.04$^{II}$ & \hskip -0.2truecm $18~36~05.3$ & \hskip -0.2truecm $-07~31~23$ & \hskip -0.2truecm $3.5$ & \hskip -0.3truecm $B$   & \hskip -0.2truecm 2,11                           & 20126+4104$^I$    & \hskip -0.2truecm $20~14~26.0$ & \hskip -0.2truecm $+41~13~32$ & \hskip -0.2truecm $1.7$ & \hskip -0.3truecm $S?$   & \hskip -0.2truecm 3,7 \\
18337-0743NE$^I$   & \hskip -0.2truecm $18~36~40.9$ & \hskip -0.2truecm $-07~39~20$ & \hskip -0.2truecm $4.0$ & \hskip -0.3truecm $BRL$   & \hskip -0.2truecm 3                         & 20216+4107$^I$    & \hskip -0.2truecm $20~23~23.8$ & \hskip -0.2truecm $+41~17~40$ & \hskip -0.2truecm $1.7$ & \hskip -0.3truecm $S$  & \hskip -0.2truecm 3.7 \\
18355-0650$^{II,*}$  & \hskip -0.2truecm $18~38~14.2$ & \hskip -0.2truecm $-06~47~47$ & \hskip -0.2truecm $4.2$ & \hskip -0.3truecm $B$   & \hskip -0.2truecm 8                   & 20319+3958$^I$    & \hskip -0.2truecm $20~33~49.3$ & \hskip -0.2truecm $+40~08~45$ & \hskip -0.2truecm $1.6$ & \hskip -0.3truecm $S$   & \hskip -0.2truecm 3,7 \\
18372-0541$^I$     & \hskip -0.2truecm $18~39~56.0$ & \hskip -0.2truecm $-05~38~49$ & \hskip -0.2truecm $1.8$ & \hskip -0.3truecm $B$   & \hskip -0.2truecm 3,7                     & 22134+5834$^I$    & \hskip -0.2truecm $22~15~09.1$ & \hskip -0.2truecm $+58~49~09$ & \hskip -0.2truecm $2.6$ & \hskip -0.3truecm $BRS$  & \hskip -0.2truecm 3,7 \\
18385-0512E$^I$    & \hskip -0.2truecm $18~41~13.3$ & \hskip -0.2truecm $-05~09~06$ & \hskip -0.2truecm $2.0$ & \hskip -0.3truecm $R$   & \hskip -0.2truecm 3,7                       & 23033+5951$^I$    & \hskip -0.2truecm $23~05~25.7$ & \hskip -0.2truecm $+60~08~08$ & \hskip -0.2truecm $3.5$ & \hskip -0.3truecm $S?$   & \hskip -0.2truecm 3,7 \\
G31.41+0.31$^{II}$ & \hskip -0.2truecm $18~47~34.7$ & \hskip -0.2truecm $-01~12~46$ & \hskip -0.2truecm $7.9$ & \hskip -0.3truecm $ ... $   & \hskip -0.2truecm 10                       & NGC7538-11$^I$    & \hskip -0.2truecm $23~13~44.7$ & \hskip -0.2truecm $+61~26~54$ & \hskip -0.2truecm $2.8$ & \hskip -0.3truecm $B$   & \hskip -0.2truecm 2 \\
18454-3$^I$        & \hskip -0.2truecm $18~47~55.9$ & \hskip -0.2truecm $-01~53~35$ & \hskip -0.2truecm $5.6$ & \hskip -0.3truecm $ ... $   & \hskip -0.2truecm 3,7                 & NGC7538-N$^{II}$  & \hskip -0.2truecm $23~13~45.4$ & \hskip -0.2truecm $+61~28~12$ & \hskip -0.2truecm $2.8$ & \hskip -0.3truecm $B$   & \hskip -0.2truecm 2 \\
18454-4$^I$        & \hskip -0.2truecm $18~48~01.4$ & \hskip -0.2truecm $-01~52~37$ & \hskip -0.2truecm $5.6$ & \hskip -0.3truecm $ ... $   & \hskip -0.2truecm 3,7                   & 23139+5939$^I$    & \hskip -0.2truecm $23~16~09.3$ & \hskip -0.2truecm $+59~55~23$ & \hskip -0.2truecm $4.8$ & \hskip -0.3truecm $BRS$   & \hskip -0.2truecm 3,7 \\
18470-0044$^I$     & \hskip -0.2truecm $18~49~36.7$ & \hskip -0.2truecm $+00~41~05$ & \hskip -0.2truecm $8.2$ & \hskip -0.3truecm $R$   & \hskip -0.2truecm 3,7                   & 23151+5912$^I$ & \hskip -0.2truecm $23~17~21.0$ & \hskip -0.2truecm$+59~28~49$ & \hskip -0.2truecm $5.7$ & \hskip -0.3truecm  $BRS?$   & \hskip -0.2truecm 3,7\\

\enddata

\tablenotetext{a}{Indices ``I'' and ``II'', attached to the source
names, denote group I and II, respectively (see Sect.\,1). ``*''
denotes the optical thin line is quoted from Luo \& Wu (2007, in
preparation); b. The positions are taken from the references (last
column). c. If there is no distance available for a core, the
distance is calculated using the galactic rotation curve of Brand \&
Blitz (1993). If there is a distance ambiguity, the nearer one is
chosen. d. Type of detected line profile (Sect.\,3.1 for
difinitions). References: 1. Tieftrunk et al. (1998); 2. Mueller et
al. (2002); 3. Beuther et al. (2002) and references therein; 4.
Zinchenko et al. (1997); 5. Molinari et al. (2000); 6. Hunter et al.
(1998); 7. Sridharan et al. (2002); 8. Wu et al. (2006) and the
references therein; 9. Hughes \& MacLeod (1994); 10. Hatchell et al.
(2000); 11. Lockman 1989.}
\end{deluxetable}

\clearpage

\begin{deluxetable}{llllll}
\tabletypesize{\small} \tablewidth{0pt} \tablecolumns{6}
\tablecaption{Blue excess statistics \label{tb3}}

\startdata \tableline \tableline
Line/Source &   $N_{\rm B}$ & $N_{\rm R}$   & $N_{\rm T}$   & $E$ & $p$ \\
\tableline
\multicolumn{6}{c}{Results from selected lines} \\
HCO$^+$ (1--0)   & 16    & 4     & 41        & 0.29  & 0.006 \\
HCO$^+$ (3--2)   & 8     & 5     & 28        & 0.11  & 0.29 \\
CS (3--2)        & 9     & 1     & 28        & $0.29$    & 0.01 \\
\multicolumn{6}{c}{HCO$^+$(1--0) line results w.r.t. source properties} \\
Group I     & 9 & 4 & 29    & 0.17  & 0.13  \\
Group II    & 7 & 0 & 12    & 0.58  & 0.008  \\
Total       & 16 & 4 & 41   & 0.29  & 0.006  \\
\enddata

\tablenotetext{\ }{$N_{\rm B}$: Number of "blue" profiles; $N_{\rm
R}$: Number of "red" profiles; $N_{\rm T}$: Total number of observed
sources. $E$: Excess; $p$: statistical likelyhood that the result is
caused by random fluctuations (see Sect.\,3.2). Note that only 15 of
the 17 inflow candidates show blue profiles in the HCO$^+$\,(1--0)
line. The other two are identified by CS\,(3--2) and HCO$^+$\,(3--2)
spectrum, respectively. One of the 16 HCO$^+$\,(1--0) blue profils
is a "BRL" source (see Sect.\,3.1).}

\end{deluxetable}

\end{document}